\documentclass[12pt,preprint]{aastex}

\usepackage{lscape}

\newcommand{\myemail}{mizuta@MPA-Garching.MPG.DE}

\slugcomment{accepted for publication in ApJ}

\shorttitle{Formation of Pillars in HII Regions}
\shortauthors{Mizuta et al.}

\begin{document}
\title{Formation of Pillars at the Boundaries between HII Regions and Molecular Clouds}

\author{Akira Mizuta\altaffilmark{1,2}, Jave O. Kane\altaffilmark{3},
Marc W. Pound\altaffilmark{4},
Bruce A. Remington\altaffilmark{3},\\
Dmitri D. Ryutov\altaffilmark{3},
and Hideaki Takabe\altaffilmark{5}}
\altaffiltext{1}{Yukawa Institute for Theoretical Physics, Kyoto
University, Oiwake-cho Kitashirakawa, Sakyo-ku, Kyoto, 606-8502 Japan.}
\altaffiltext{2}{Max-Planck-Institute f\"ur Astrophysik
Karl-Schwarzschild-Str. 1, 85741 Garching, Germany, E-mail:\myemail}
\altaffiltext{3}{University of California, Lawrence Livermore National
Laboratory, 7000 East Ave., Livermore, CA 94551 U.S.A.}
\altaffiltext{4}{Department of Astronomy, University of Maryland,
College Park, MD 20742 U.S.A.}
\altaffiltext{5}{Institute of Laser Engineering Osaka University,
2-6 Yamada Oka, Suita, Osaka, 565-0871, Japan}

\begin{abstract}
We investigate numerically the hydrodynamic instability of an
ionization front (IF) accelerating into a molecular cloud, with
imposed initial perturbations of different amplitudes.  When the
initial amplitude is small, the imposed perturbation is completely
stabilized and does not grow.
When the initial perturbation amplitude is large enough,
roughly the ratio of the initial amplitude
to wavelength is greater than 0.02,
portions of the IF temporarily separate from the molecular cloud surface,
locally decreasing the ablation pressure.
This causes the appearance of a large, warm HI region and
triggers nonlinear dynamics of the IF.
The local difference of the ablation pressure and acceleration
enhances the appearance and growth of a multimode perturbation.
The stabilization usually seen at the IF in the linear regimes
does not work due to the mismatch of the modes of the perturbations
at the cloud surface and in density in HII region
above the cloud surface.
Molecular pillars are observed in the late stages of the
large amplitude perturbation case.  The velocity gradient in the
pillars is in reasonably good agreement with
that observed in the Eagle Nebula.
The initial perturbation is imposed in three different ways: in
density, in incident photon number flux, and in the surface shape.
All cases show both stabilization for a small initial perturbation
and large growth of the second harmonic by increasing amplitude
of the initial perturbation above a critical value.

\end{abstract}


\section{INTRODUCTION}
The shapes of the surfaces between molecular clouds and the HII
regions around massive stars typically feature elongated structures,
commonly referred to as pillars, columns, spikes, or elephant
trunks \citep{Hester96,Pound98,Pound03,Bally03}.  The surface is
a photoionization front driven by the strong UV radiation from
the OB stars.  One well-known example is the Eagle Nebula which
has three large, molecular pillars near a small group of O stars
\citep{Hillenbrand1993}.  Although a number of theoretical and
numerical studies have been done, the formation mechanism is still
not fully understood.

The OB stars are hot, with photospheres at temperature of a few
tens of thousands of Kelvins \citep{Dors03}, and give off high
intensity UV photons.
These UV photons irradiate the molecular
cloud which surrounds the OB stars and photoevaporation occurs,
resulting in a stratified structure.
The photoevaporative flow
(ablated plasma) velocity is normal to the molecular surface \citep{Hester96}.
The region between the OB stars and the
molecular cloud surface is the HII region in which the hydrogen gas is
almost fully ionized and the photoionization and recombination
to neutral atomic hydrogen occurs in steady state.  The IF is
a very thin layer because of the short mean free path for the
incident photons on the cloud surface.  At this cloud surface
all the photons above the Lyman limit are absorbed.
The photons below the ionization limit but above 11.2 eV cannot ionize
the hydrogen atom  in the ground state but can penetrate the IF
and dissociate the molecular hydrogen in the underlying layers.
The thickness of the dissociation front differs with each cloud.
The last layer in the stratified structure is the molecular gas.
Because of strong radiative cooling, the molecular cloud temperature
is typically a few tens of Kelvin  \citep{Neufeld95}.  The dynamics
of such irradiated molecular clouds is thought to play a role in star
formation \citep{Bally03,Sugitani02,MaCaughrean02}.  Consequently,
the outflow dynamics from the molecular cloud into the HII regions
is of considerable, general interest \citep{Pound03,Mizuta02}.

\begin{figure}
\begin{center}
\epsscale{.80}
%\plotone{fig2_1.eps}
%$\rotatebox{270}{\plotone{m16LocalStarsBW.ps}}
\rotatebox{270}{\plotone{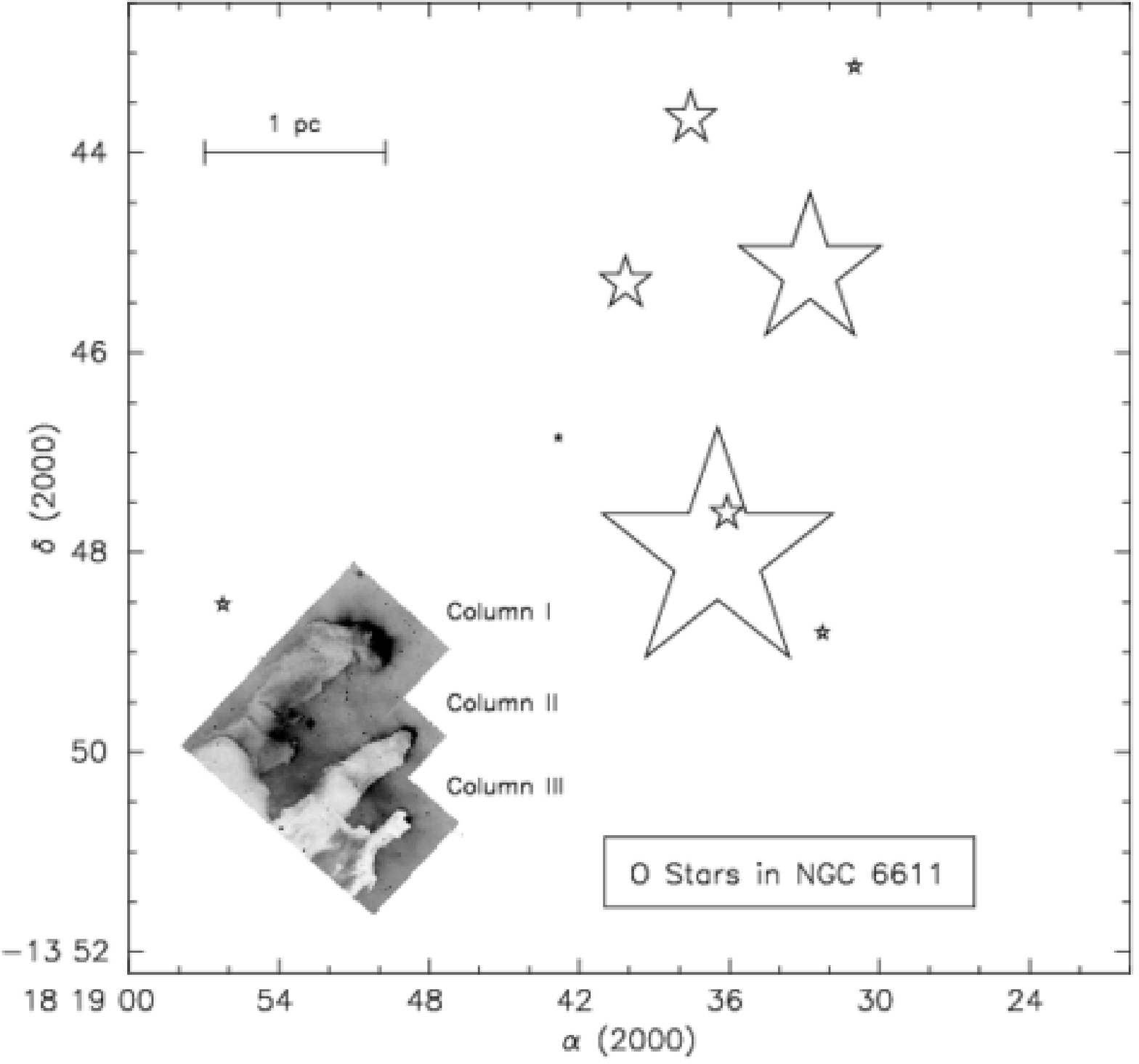}}
\caption{Schematic diagram showing the position of the O stars that are
photoevaporating the Eagle Nebula molecular cloud
taken from Fig. 1 in \citet{Pound98}.
The size of the symbols indicates the relative strength of the Lyman
continuum flux from each star.
Most of the ionizing radiation comes from two stars of spectral type O5
V and O5.5 V in the nearby young cluster NGC 6611.}
\label{poundEagle}
\end{center}
\end{figure}

The pillars in M16, popularly known as the Eagle Nebula, are
at a distance of $\sim 2$ kpc (\citep{Hillenbrand1993}).  It is
estimated that half of the ionizing radiation comes from a single
O3-4 star and the rest mostly comes from the three other nearby
stars (O5-6) \citep{Hester96}.
The orientation of this system
can be seen in Fig. \ref{poundEagle} (taken from \cite{Pound98}).
The total ionizing flux is estimated to be $S\sim 1.2\times10^{50} \mbox{
s}^{-1}$.  The temperature in the HII region near the molecular
cloud surface is 9500~K and the ionized hydrogen number density
is $5~\times~10^{3} \mbox{ cm}^{-3\  }$ \citep{Levenson2000}.
The thickness of the photodissociation region (PDR) is relatively thin
because the large hydrogen number density ($\mbox{n(H)}\sim 10^{3}
\mbox{ cm}^{-3\ }$) provides high optical depth.  As measured by
CO line observations, the velocity gradients along the long axes
of the pillars (from ``head'' to ``tail'') are between -20.7 and
+6.7 $\mbox{km s}^{-1} \mbox{pc}^{-1}$, with an average magnitude
of 8.3 $\mbox{km s}^{-1} \mbox{pc}^{-1}$.  Furthermore, the pillars
are not in the same plane in the sky; the differing signs of the
velocity gradients indicate inclination towards (positive gradient)
or away (negative gradient) from the observer.
Predictions of
the classical Rayleigh-Taylor (RT) instability theory with regard
to the velocity gradient \citep{Frieman54} are incompatible with
the observed gradient, at least for constant acceleration
\citep{Pound98}.
Another formation mechanism should be considered.

The formation mechanism of pillars beside massive stars has
been discussed by a number of authors.  There are two classes
of models for their formation.  One is that the pillars are
formed due to dense, pre-existing cores in the molecular
cloud (e.g. \citet{Reipurth1983}; \citet{Bertoldi1990};
\citet{Lefloch1994}; \citet{Williams01}).  In a interesting
variation on this idea, \citet{Williams01} also simulated the result
of a radiatively driven, pre-existing, short pillar with the same
density as the background molecular cloud (their Case II).  Even in
this case, the pillar grows.

Another possibility is that the pillars are caused by hydrodynamic
instabilities.  \citet{Spitzer54} proposed a model based on the RT
instability occurring at the contact discontinuity between the dense
molecular cloud and the lower density, hot, photoevaporated plasma.
He suggested that the pillars in the Eagle Nebula were the result
from the nonlinear stage of the RT instability.  \citet{Frieman54}
estimated the time scale of the instability to be less than $10^6$
years.
\citet{Pottasch58} showed observations of several
nebulae and estimated the age of the pillars by the RT model, assuming
small initial amplitude of the perturbations.
\citet{Vandervoort62} found another type of instability
at the IF in a non-accelerating
frame, the so-called IF instability, which, in his analysis,
was present in the case of non-normal incident radiation.
He derived a dispersion relation for perturbations growing by this process.
\citet{Axford64} extended it to include recombination,
which plays a crucial role in the HII region.  The recombination
in the ionized gas works to smooth the surface when the wavelength
of the perturbation is much larger than the recombination length.

\citet{Williams01} showed robust development of photoionized pillars 
with an isothermal model, although they considered a semi-infinite
cloud and a non-accelerating IF. \citet{Williams02} derived a
dispersion relation for the IF with a non-normal incident radiation
field including the effect of recombination in the ionized gas.
\citet{Ryutov03} found unstable modes with non-normal radiation
for an accelerating IF, but did not include recombination.

Most studies have assumed a semi-infinite molecular cloud and
non-accelerating IF.  Instabilities at an accelerating IF were
studied numerically by Mizuta et al. (2005, hereafter Paper I).
They show that, for imposed perturbations with small initial
amplitude, i.e. in the linear regime, there was no significant
growth.  Large growth of a classical RT instability is observed
when recombination in the HII region is not included.  We study
here in more depth the dynamics of an accelerating IF.
The existence of the acceleration at the IF is quite different
than most previous models.
This paper is organized as follows.
The numerical method and conditions are described in Sec. \ref{model}.
The results and discussions are given is Sec. \ref{discussion},
and the conclusion is given in Sec \ref{conc}.

\section{MODEL : ACCELERATING IF}
\label{model}
The same physics and computational method used in Paper I are
included in this study.  The energy balance and magnetic pressure to
prevent radiative collapse for the molecular cloud are considered.
The equations we numerically solve are:
\begin{eqnarray}
{\partial \rho \over \partial t}+
\nabla\cdot (\rho\mbox{\boldmath$u$})=
0, \label{e-massconserv} \\
%%%%%%%%%%%%%%%%%%%%%%%%%%%%%%%%%%%%%%%%%%%%%%%
{\partial (\rho\mbox{\boldmath$u$})\over \partial t}+
\nabla\cdot (\rho\mbox{\boldmath$u$}\mbox{\boldmath$u$}+p\mbox{\boldmath$I$})=
0, \label{e-momentumconserv} \\
%%%%%%%%%%%%%%%%%%%%%%%%%%%%%%%%%%%%%%%%%%%%%%%
{\partial \over \partial t}\left(\rho\left({1\over 2}\mbox{\boldmath$u$}^2
+\epsilon\right)\right)+
\nabla\cdot 
\left(
\left(\rho\left({1\over 2}\mbox{\boldmath$u$}^2
+\epsilon\right)+p\right)\mbox{\boldmath$u$}
\right)\nonumber \\
=-q_{re}+q_{uv}-q_{mol}, \label{e-energyconserv}\\
%%%%%%%%%%%%%%%%%%%%%%%%%%%%%%%%%%%%%%%%%%%%%%%
p={2(3f+1)\over 7f+5}\rho \epsilon +p_{_{M}}
\left({\rho\over \rho_{_{M}}}\right)
^{\gamma_{_{M}}}, \label{e-eos} \\
%%%%%%%%%%%%%%%%%%%%%%%%%%%%%%%%%%%%%%%%%%%%%%
n{\partial f\over \partial t}+
n \mbox{\boldmath$u$}\cdot\nabla f=
an(1-f)|\mbox{\boldmath$J$}|-\alpha_{B}n^2f^2,\label{e-fevolution}\\
%%%%%%%%%%%%%%%%%%%%%%%%%%%%%%%%%%%%%%%%%%%%%
\nabla\cdot\mbox{\boldmath$J$}
=
-an(1-f)|\mbox{\boldmath$J$}|
,\label{e-radtransport}
\end{eqnarray}
where $\rho$ is mass density, $p$ is pressure,
$\mbox{\boldmath$I$}$ is unit tensor,
\mbox{\boldmath$u$} is the velocity vector,
and $\epsilon$ is the specific internal energy.
The equations (\ref{e-massconserv})-(\ref{e-energyconserv})
are mass, momentum and energy conservation
with energy sources.
The energy source terms due to
recombination in the ionized region, absorption of the UV radiation
from OB stars, and cooling in the molecular gas are $q_{re}, q_{uv},
q_{mol}$, respectively.
These energy sources are evaluated as
\begin{eqnarray*} 
q_{re}  & = & (nf)^2~\beta_{B}~k~T, \\
%%%%%%%%%%%%%%%%%%%%%%%%%%%%%%%%%%%%%%%%%%%%%%%
%
q_{uv}  & = & W a n(1-f)|\mbox{\boldmath$J$}|, \\
%%%%%%%%%%%%%%%%%%%%%%%%%%%%%%%%%%%%%%%%%%%%%%%
%
q_{mol} & = & n_{mol}^2\times 
10 ^{-29}~\mbox{erg cm}^{-3}\mbox{s}^{-1}, \\
%%%%%%%%%%%%%%%%%%%%%%%%%%%%%%%%%%%%%%%%%%%%%%%
%
T & = &(m_{p}/k)\times[4\epsilon / (7f+5)], \\
%%%%%%%%%%%%%%%%%%%%%%%%%%%%%%%%%%%%%%%%%%%%%%%
%
n_{mol} & = & n~(1-f)/2,  \\
%%%%%%%%%%%%%%%%%%%%%%%%%%%%%%%%%%%%%%%%%%%%%%%
%
%\rho \epsilon & = & 2\times1.5 (nf) k T + 2.5 n_{mol} k T,
\rho \epsilon & = & 3 (nf) k T + 2.5 n_{mol} k T,
%%%%%%%%%%%%%%%%%%%%%%%%%%%%%%%%%%%%%%%%%%%%%%%
\end{eqnarray*}

\noindent where $T$ is the temperature in Kelvins
$m_p$ is the proton mass,
$k$ is the Boltzmann constant,
and $n$ and $n_{mol}$ are atomic hydrogen and molecular hydrogen number
density.
We ignore metal line cooling for simplicity, since its cooling
power has the same dependence as that of recombination (proportional
to ionized hydrogen number density).
The dissociation heating in the neutral region 
and other radiative
processes are also ignored for simplicity.
The heating function is $W=1.73 \times 10^{-12}$ erg
which corresponds to the average energy deposited
into the gas per absorbed ionizing photon,
and leads to produce an isothermal temperature of $T=10^4$ K
in the ionized gas as an
equilibrium state of photoionization heating and recombination
cooling.
The case B recombination coefficients,
which are summation of all recombination coefficients
of hydrogen except the recombination to the ground state,
are assumed constant
$\alpha _B=2.6\times10^{-13}\mbox{cm}^{3}\mbox{s}^{-1}$ and
$\beta_{B}=1.25\alpha_{B}$ (at $T=10^4$ K from \citet{Hummer63}),
where $\beta_{B}$ includes 
the thermal velocity dependence of the rates of recombination and
free-free collisional cooling.
The equation of state (Eq. \ref{e-eos}) includes a magnetic pressure
term for dense gas to prevent radiative collapse \citep{Ryutov02},
where $f\equiv n_{i}/n$ is the ionization fraction, $n_{i}$ is the
ionized hydrogen volume density, $p_{_{M}}$ and $\rho_{_{M}}$ are
constant values.  The index $\gamma_{M}$ ($=4/3 $) is also constant.
Numerically, we do not take into account the atomic hydrogen state,
assuming the dissociation front is thin as in the Eagle nebula.
Equations (\ref{e-fevolution}) and (\ref{e-radtransport}) describe
the evolution of $f$ and the transport of the incident radiation, where
$a=6\times 10^{-18}\mbox{cm}^{2}$ is the photoionization cross-section
of hydrogen, and $\mbox{\boldmath$J$}$ is the number flux of
ionizing photons, i.e., photons $\mbox{cm}^{-2} \mbox{s}^{-1}$.
Since we use the ``on-the-spot approximation''
(i.e., photons emitted in recombination to ground state
are immediately reabsorbed,
whereas the photons emitted in recombination to
second or higher levels are assumed to escape from the system),
we do not consider the diffusive photon emission and transport.

The 2D computational domain is ($x\times y$) = (0.46 pc$\times$3 pc).
Uniform grid points ($\Delta x = \Delta y = 2.5\times 10^{-3}$
pc) are used.  Periodic boundary conditions are employed 
($x=0$ and $x=0.46$ pc).
The other boundary conditions are open boundary
which means zero gradient.
An incident photon flux
($\mbox{\boldmath $J_0$} =
-5\times 10^{11} {\hat y} \mbox{ cm}^{-2}\mbox{s}^{-1}$)
is imposed from one boundary ($y=3$ pc), where ${\hat y}$ is the
unit vector of the $y$ axis.
The rays are parallel to the $y$
axis.  A cloud of initial thickness of a quarter pc, and density
of $n(H_2)=10^{5} \mbox{ cm}^{-3}$ is set 0.5 pc away from the
boundary where the incident photons come in.
A very dilute gas $n(H)=10 \mbox{ cm}^{-3}$ is imposed in the other regions.
Initially, the gas at $y>2.25$ pc is in pressure
equilibrium, and the gas at $y<2.5$ pc is isothermal (40 K).
The constant parameters in the equation of state, such as,
$\rho_{_{M}}$ and $p_{_{M}}$, are 
the mass density and thermal pressure,
of the initial molecular cloud, respectively.

We impose the perturbation in the cloud
in three different ways.  The first is a
density perturbation in a layer of 0.125~pc thickness along the surface
of the cloud, with the form $n=n_0(1-A\cos [2\pi x/0.46\mbox{ pc}])$,
where $A$ is the amplitude of the perturbation and $n_0=10^{5}
\mbox{ cm}^{-3}$.  We assume a wavelength of $\lambda=0.46$ pc,
and consider amplitudes $A=0.2$ (model D2), 0.3 (D3), 0.4 (D4)
and 0.5 (D5).  The second method is a 30\% amplitude perturbation
in the incident photon number flux of form $\mbox{\boldmath$J$}=
\mbox{\boldmath$J_0$}(1-0.3 \cos[2\pi x/0.46\mbox{ pc}])$, starting
at 98, 102, 106, or 108 kyr and ending at 110 kyr.  These are models
P098, P102, P106, P108, respectively.  This is similar to the method
used in Paper I and in Mizuta et al. (2005b).  The third method is
to impose surface perturbations on the initial cloud, according
to $y=2.5-C\cos[2\pi x/0.46\mbox{ pc}]$.  The amplitude ($C$) is
$3.8\times10^{-3}$ pc (model S038), $9.0\times10^{-3}$ pc (S090),
$1.4\times10^{-2}$ pc (S140), and $1.9\times10^{-2}$ pc (S190).

\section{RESULTS and DISCUSSION}
\label{discussion}
\subsection{1D Dynamics}
\label{1D}
\begin{figure}
\epsscale{.80}
%\plotone{normalrho.eps}
\plotone{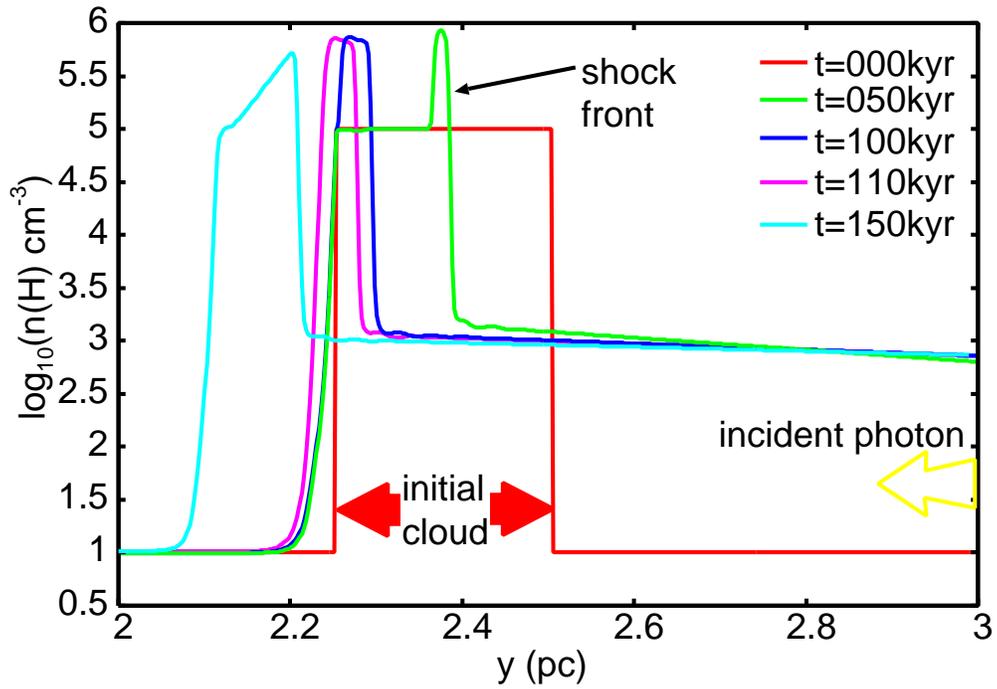}
\caption{ 
One dimensional hydrogen number density profiles of the case without
perturbations at five different times, $t=0, 50, 100, 110$, and
$150$ kyr.  The profiles at three phases of the dynamics,
compression ($0\le t \lesssim 100$ kyr), rarefaction 
($100 \lesssim t \lesssim 110$ kyr), and acceleration 
($t \gtrsim 110$ kyr) are shown.}
\label{1D_1}
\end{figure}

We briefly show the results without any perturbations,
before showing the results with perturbations.
The dynamics
without any perturbation is very simple.  When the incident photon
flux is turned on, a shock propagates through the molecular cloud
(compression phase).
After this shock breaks out of the back side
of the cloud, a rarefaction passes back through the shocked cloud.
Then an acceleration phase begins at about $t=100$ kyr, as the
cloud moves as a unit.
Figure \ref{1D_1} shows one dimensional
profiles of the hydrogen number density at
the early phase of the dynamics at $t=0, 50, 100, 110,$ and 150 kyr.
In the figure, all three phases are shown, although the phase at
around $t=100$ kyr to $t=110$ kyr in which the rarefaction passes
back through the compressed cloud is too short to show.

\begin{figure}
\epsscale{.80}
%\plotone{compare.eps}
\plotone{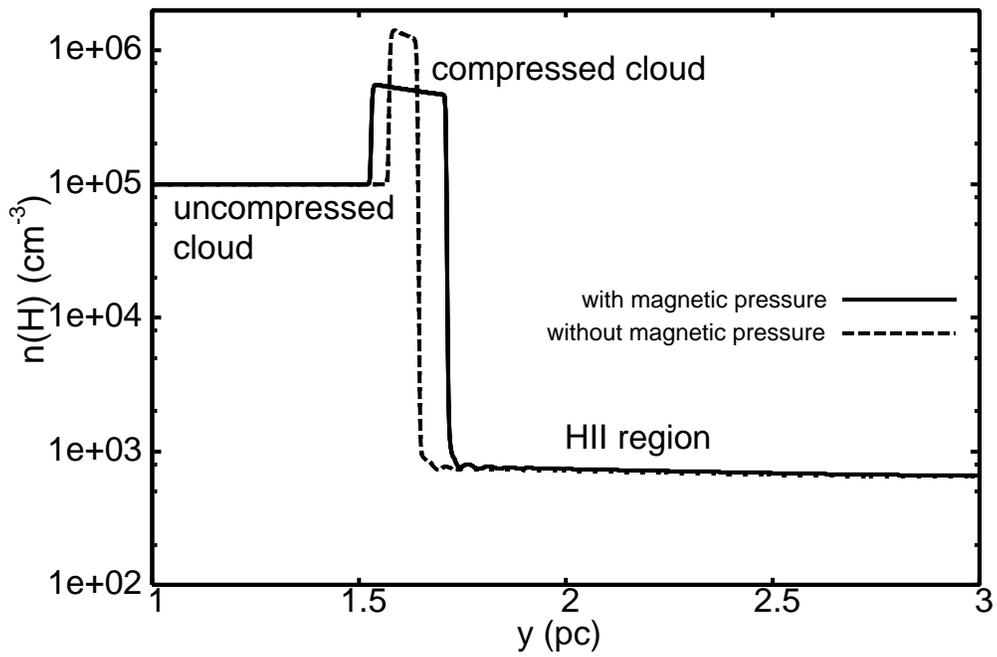}
\caption{ 
One dimensional hydrogen number density profiles of the initial
semi-infinite cloud.  The cases with and without magnetic pressure
at $t=500$ kyr are shown.  The number density of the compressed cloud
without magnetic pressure is about 1.5 times higher than that
with magnetic pressure.
}
\label{1D_2}
\end{figure}

The hydrogen number density profiles of an initially semi-infinite
cloud are shown in Fig. \ref{1D_2}, both with and without magnetic
pressure.  The initial total pressure without the magnetic component
is lower than with it.  Without magnetic pressure, the number
density of the compressed cloud becomes about 1.5 times higher,
since there is less pressure support.

\subsection{Small Initial Amplitude Cases}
\label{small}
\begin{figure}
\epsscale{.50}
%\plotone{eagle_paper_h.eps}
%\plotone{eagle_paper3.eps}
\plotone{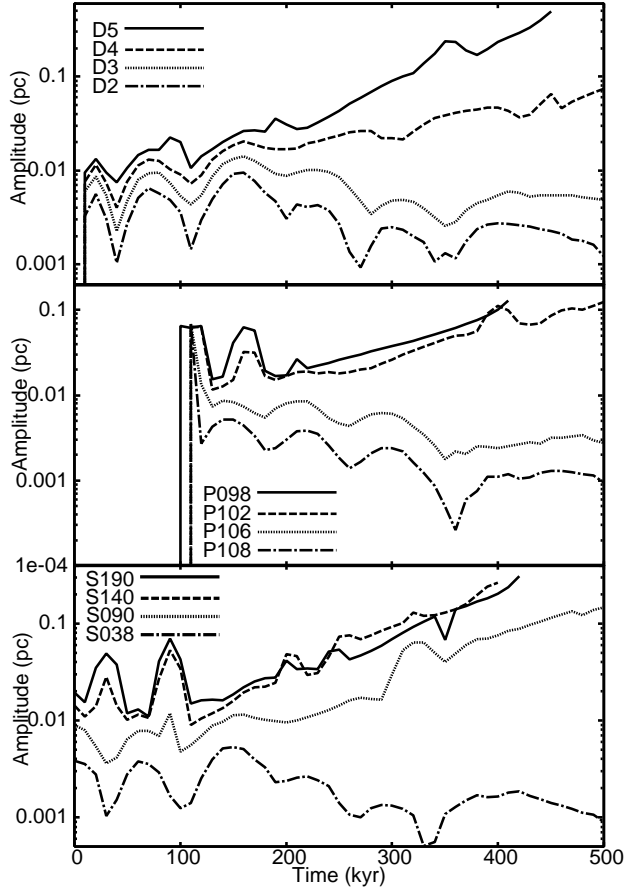}
%\plotone{fig1.eps}
\caption{
The time evolution of the amplitude for different methods of
imposing the initial perturbation.  Top: the initial perturbation
is imposed in density.  Middle: the perturbation is imposed in the
photon number flux.  Bottom: a surface perturbation is imposed.
Note the amplitude is defined as half the peak-to-valley height
of the perturbation in the $f=0.5$ contour in the $y$ direction.
This does not always corresponds to the amplitude of the perturbation
on the surface of the molecular cloud due to the ``separation
effect'' (see text).  When the imposed initial perturbation becomes
large enough, the perturbations grow in all cases.  When the initial
perturbation is small, the growth is stabilized by recombination.
}
\label{amp}
\end{figure}

Figure \ref{amp} shows the evolution of the amplitude, i.e.,
half the peak-to-valley of the perturbation of the $f=0.5$
contour in the $y$ direction for each case, for all three methods
of imposing the perturbation.  It should be noted that this does not
always correspond to the amplitude of the perturbation on the cloud
surface because of the effect of ``separation of the IF'' discussed
below.  In cases with small initial perturbations (models D2, D3,
P106, P108, and S038), the perturbation does not grow  but rather
oscillates as observed in Paper I.  This stabilization is caused by
the density profile in the HII region.  The ablated (photoevaporated)
gas in the concave region of the incipient bubble concentrates
(gets weakly ``focused'') in the region above the bubble vertex.
This concentration of gas subsequently absorbs more of the incoming
ionizing photons, reducing the photon flux reaching the IF region
at the cloud surface near the bubble vertex, compared with the
spike region.  This is because the recombination rate depends on the
square of the ionized hydrogen number density in the HII region.
The net result is that the ablation pressure locally decreases
in the bubble region compared to the spike region, which acts to
smooth out the perturbed surface, if the perturbation is small.
This is also shown in Fig.\ref{density}a with a 2D plot of number
density late in time ($t=460$ kyr) for a small initial density
perturbation (model D2).  The IF is quite smooth, namely, no growth
of the perturbation has occurred.  The other cases (D3, P106, P108,
and S038) in which the perturbation does not grow behave similarly.

\subsection{Large Initial Amplitude Cases}
When the perturbation initial amplitude is larger,
roughly when the ratio of the initial amplitude to wavelength
is greater than 0.02, the perturbations
grow, after the acceleration phase begins at around $t=100$ kyr
(models D4, D5, P098, P102, S090, S140, and S190).  Recombination in
the HII region is included in all cases.  Here, we concentrate on the
case with an imposed density perturbation (model D5) to illustrate
why the perturbation grows when the amplitude of the imposed
perturbation exceeds a critical value.

\begin{figure}
\epsscale{.88}
%\plotone{density_panel.eps}
\plotone{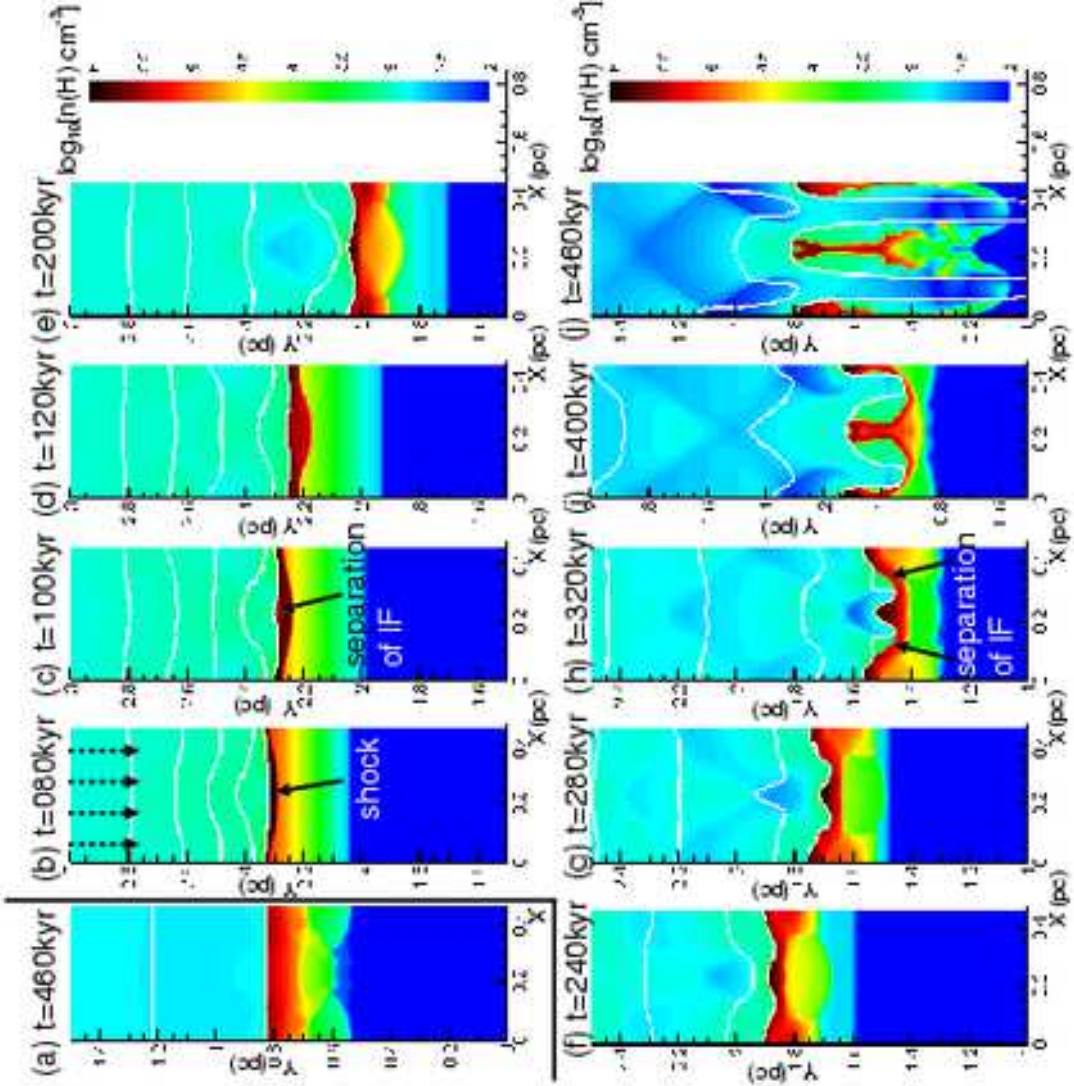}
%\plotone{fig2.eps}
\caption{
Log density (color) and incident photon flux contours (white
solid curves) from 0 to $|\mbox{\boldmath$J$}|=5\times 10^{11}
\mbox{ cm}^{-2}\mbox{s}^{-1}$ with uniform intervals of
$|\mbox{\boldmath$J$}|=1\times 10^{11} \mbox{ cm}^{-2}\mbox{s}^{-1}$.
The incident photon flux, which is parallel to the $y$ axis,
comes in from the top boundary.
The dashed arrows
in (b) indicate the initial photon flux from the top boundary.
(a) A 30\% initial perturbation
in density is imposed (model D3), and no growth has occurred even
at late times ($t=460$ kyr).  (b-j) Similar to (a) except than
the initial density perturbation is 50\% (model D5).  The times
corresponding to each 2D density snapshot are indicated at the top
of each plot.  (b) A shock is passing through the cloud, as indicated
by the arrow.  Note the $|\mbox{\boldmath$J$}|=1\times 10^{11}
\mbox{cm}^{-2}\mbox{s}^{-1}$ contour has the opposite phase as the IF.
(c-g) The shock breaks out the back of the molecular cloud, and
the acceleration phase begins.  
The solid arrows in (c) and (h) indicate the
IF separation from the high density molecular cloud ``surface''.
(h)-(i) Due to recombination, the IF separates from the high-density
molecular cloud surface, which locally lowers the ablation pressure,
and in (i) allows the perturbation spike in the center to grow due
to the local pressure difference.  (j) Once the perturbation growth
exceeds a critical amplitude, the stabilization effect is reduced,
and the evolution quickly enters the deep nonlinear regime, creating
a pillar.
}
\label{density}
\end{figure}

Figures \ref{density} (b)-(i) show a series of the 2D hydrogen
number density plots (color) with incident photon flux contours
(solid curves) at different times.  The high density regions
correspond to molecular gas compressed by the ablation pressure.
A typical hydrogen number density in the HII region near the IF
is $10^3 \mbox{ cm}^{-3}$ as shown in Figure \ref{1D_1}.
The solid white lines correspond to incident photon number contours from
0 to $5\times10^{11}\mbox{ cm}^{-2}\mbox{s}^{-1}$ at intervals of
$1\times10^{11}\mbox{ cm}^{-2}\mbox{s}^{-1}$, starting from the
ionization front where the ionization fraction ($f$) goes
to zero.  Since we do not show the whole computational domain,
all the photon flux contours are not shown in the figures in the
later phases.

\begin{figure}
\epsscale{.80}
%\plotone{zoomin.eps}
\plotone{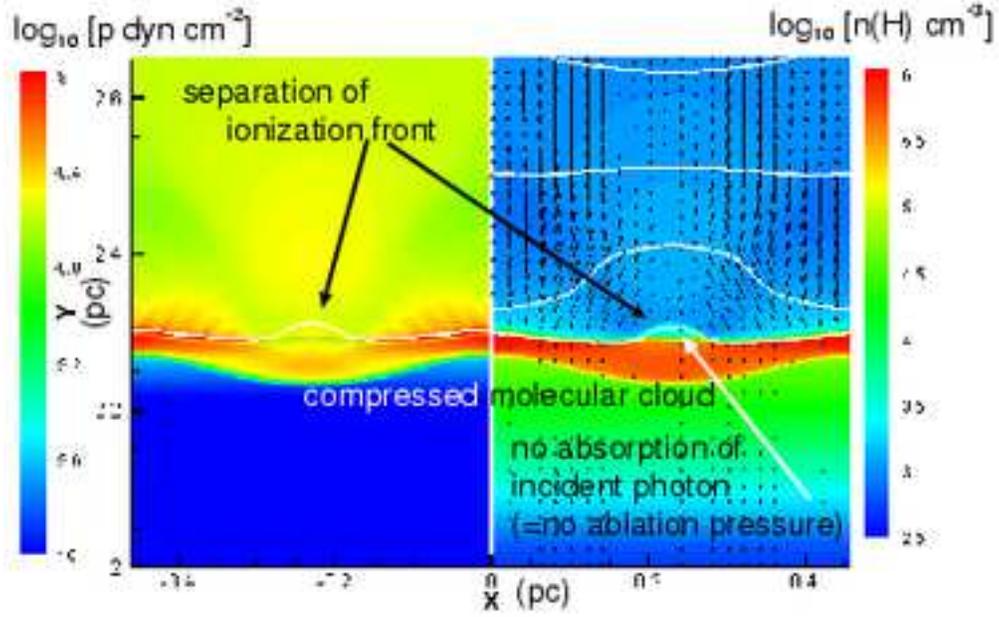}
%\plotone{fig3.eps}
\caption{
A close up of Fig. \ref{density}(c) around the IF.  Density
($x>0$) and pressure ($x<0$) are plotted (color).  The contours of
the incident photon flux ($|\mbox{\boldmath$J$}|=0, 1\times 10^{11},
2\times  10^{11}, 3\times 10^{11}, 4\times 10^{11} \mbox{
cm}^{-2}\mbox{s}^{-1}$) are plotted ($x>0$).  The contour of the
ionization degree ($f=0.5$) is plotted ($x<0$).
The arrows in the right panel are velocity field.
The IF slightly
separated from the cloud surface at around $x=0.23$, $y=2.3$ pc,
since all the incident photons are absorbed by the recombined
hydrogen in the high density region (around $x=0.23, y=2.4$).
A slight expansion of the cloud in the $y$ direction can be seen.
} 
\label{zoom} 
\end{figure}

In Fig.\ref{density} (b), the photon flux contours near to the IF
are of the opposite phase as the IF because of the oscillation of
the IF.  This oscillation occurs by the stabilization mechanism
seen in the linear regime (Paper I).  In Fig.\ref{density} (c),
the $|\mbox{\boldmath$J$}|=0$ photon flux contour separates slightly
from the highly compressed molecular cloud around $x=0.23$ pc,
$y=2.3$ pc (see also the close-up view in figure \ref{zoom}).
This is the separation of the IF
from the cloud surface.  It happens because all the incident photons
moving toward this region are absorbed by the recombined neutral hydrogen
accumulated near the bubble vertex at $x=0.23$ pc.  This could be
understood as an extreme case of the normal stabilization mechanism
of the IF instability, because the increasing density around the
cavity causes strong absorption of the incident photons.
During the separation of the IF, the cloud surface is shadowed.
This shadowed region does not feel any ablation pressure because no
absorption of the incident photons occur there, whereas the other
parts of the cloud surface which are directly ablated by the incident
photons feels strong ablation pressure.

The region between the shadowed cloud surface and the separated IF becomes
neutral.  The appearance of the neutral region happens because the
timescale for recombination is roughly $(\alpha_B n_i)^{-1}$
$\sim 10^2$ yr for ionized hydrogen at a number density of
$n_i=10^3\mbox{cm}^{-3}$.
Thus, once the cloud surface is shadowed, the neutral
region quickly appears close to the shadowed surface.
We can identify this region as a warm HI region ($T\sim 4000$ K )
because it has lost half of its thermal energy by recombination cooling.
The pressure in this warm HI region is a few times smaller than that
of the HII region and the molecular gas.  This region has
similar physical properties as the warm HI layer in front of PDRs
described by \citet{Hollenbach97} (see their Figure 3), which
also arises due to a lowering of the ionization fraction.

The appearance of this separation strongly affects the dynamics
of the IF.  In reaction, the compressed cloud expands into
the low pressure region, namely, in the $y$ direction, which
triggers the second harmonic of the imposed initial perturbation
in subsequent frames (Fig.\ref{density}(d) - (e)).  Although the
second harmonic of the imposed initial perturbation of the cloud
surface can be clearly seen at $t=200$ kyr, the second photon contour
($|\mbox{\boldmath$J$}|=1\times10^{11}\mbox{ cm}^{-2}~\mbox{s}^{-1}$)
from the IF still shows just a single mode.  In other words, the density
around the side ($x=0, 0.46$ pc) is high, and the density around
the center ($x=0.23$) is low, although both regions on the cloud surface
are spikes.
Thus the stabilization mechanism seen in the linear regime (Paper
I, and section ~\ref{small}) does not work in this case.  The ablation
pressure is strong around the center ($x=0.23$ pc), because the two
last contours are close to each other there
(meaning the photon flux incident on the cloud surface is relatively
high).
As a result, a large
amount of thermal energy is deposited there and compression
by the strong ablation pressure occurs.  By contrast, around the
side ($x=0, 0.46$ pc) the last two incident photon contours are
separated from each other
(meaning a lower photon flux reaches the ablation front at the cloud
surface).
The resulting lower ablation pressure causes
slight local expansion in the $y$ direction (Fig.\ref{density} (e)
and (f)).  This expansion allows the growth of the spike at the sides
(Fig.\ref{density} (g)), and a strong second harmonic has grown up.

By 320 kyr, separation of the IF from the ablation front (cloud surface)
near the lateral boundaries has occurred (Fig.\ref{density}(h)), and by 400
kyr (Fig.\ref{density}(i)), this separation has also appeared near
the middle.  The net result is that the perturbation amplitudes at
the center and lateral edges grow.  Once a critical amplitude is
exceeded, at around 400 kyr, the stabilization mechanism is weakened
by the steepness of the perturbations
(i.e., the photon flux and therefore ablation rate along the steep sides
of the perturbations drops, so there is much less density rise in the
HII gas in the concave region),
and the RT evolution proceeds
rapidly into the deep nonlinear regime (Fig.\ref{density}(j)).

\begin{figure}
\epsscale{.50}
%\plotone{resolution.eps}
%\plotone{fig4.eps}
\plotone{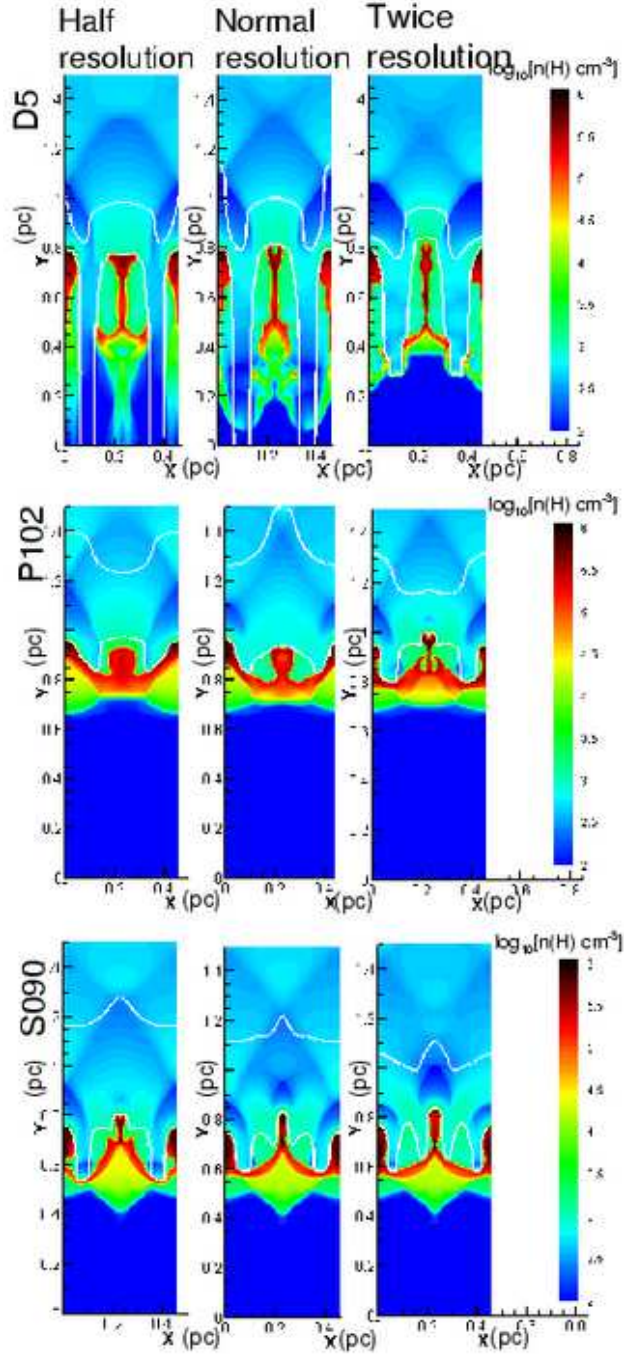}
\caption{
Lower (left panels: half the number of grid points in $x$),
standard zoning (middle panels)
and higher (right panels: twice the number of grid points in $x$)
resolution calculations.
The contours show the same physical values as in
Fig.\ref{density}.
The top panels are model D5 at $t=460$ kyr,
the middle
panels are model 102 at $t=440$ kyr,
and the bottom panels are S090
at $t=460$ kyr.
All the cases show large growth of the second
harmonic of the imposed initial perturbation, differing only in
the fine-scale structure.
}
\label{resolution}
\end{figure}

For the other cases with a large amplitude of initial perturbations
(D4, P098, P102, S090, S140, and S190), we observe similar dynamics
of the IF and cloud surface.  The separation of the IF occurs, a
second harmonic of the imposed perturbation is triggered, and finally
two pillars grow.  The center panels of Fig. \ref{resolution} show a
large perturbation growth in later phase of the dynamics using the
different methods of the imposing the initial perturbation :
density modulation (model D5; top),
perturbation in photon number flux (model P102;
middle), and surface perturbation (model S090; bottom).  The growth
of the second harmonic mode of the perturbation is observed in all cases.

\subsection{Velocity Gradient}
\begin{figure}
\epsscale{.80}
%\plotone{vel3.eps}
%\plotone{vel4.eps}
\plotone{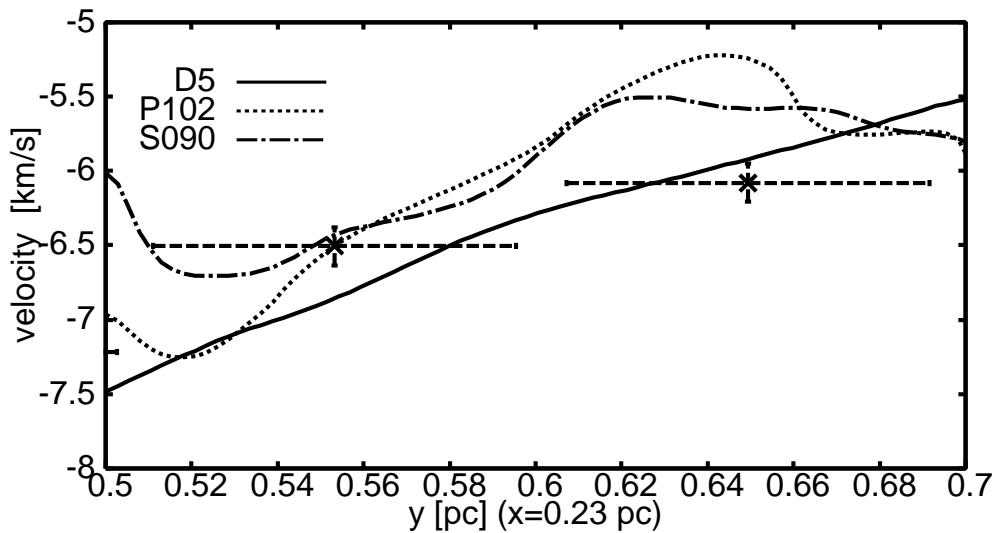}
%\plotone{fig5.eps}
\caption{
The velocity profile in the $y$ direction at $x=0.23$ pc, for
the different methods of initial perturbation: D5 ($t=460$ kyr), P102
($t=480$ kyr), and S090 ($t=480$ kyr).  The pillar is oriented such
that the right side represents its top and the left side represents
its bottom.  The points with error bars show
the measured radial gas velocities 
in Pillar II of the Eagle Nebula from \citet{Pound98},
divided by $\sin(15)$
to account for the line-of-sight inclination.
The velocity gradients
of models are in good agreement with the observed gradient.
}
\label{velocity}
\end{figure}

Figure \ref{velocity} shows the $y$-component of the velocity in
the pillar for an initial perturbation in density (model D5), in photon flux
contour (P102), and in the surface (S090).  The $V_y$ gradients
are $\sim$ 8.0 (D5), 16 (P102), and 12 (S090) $\mbox{km s}^{-1}
{\mbox {pc}}^{-1}$.
In a real pillar, an observer would see any velocity gradient
corresponding to $V_y\sin(i)$, where
$i$ is the inclination angle of the pillar with the plane of the sky.
Indeed, \citet{Pound98} measured such gradients in the pillars of
the Eagle Nebula.
This observed velocities in Pillar II of the Eagle Nebula are also
plotted for comparison. For the purposes of the figure, we have
taken $i = 15$ degrees, dividing the observed values by $\sin(15)$.
Thus, for small inclination angles, our numerical results are good
agreement with the observed Pillar II velocity gradient of $2.2
\mbox{ km s}^{-1} {\mbox {pc}}^{-1}$ (\citet{Pound98}).

\subsection{Resolution Effects}
The calculation without the recombination effect, which was shown
in Paper I, shows a large growth of the imposed perturbation.
The surface also shows growth of a shorter wavelength perturbation
due to purely numerical reasons, such as less dissipation.
It is impossible to impose exactly one single mode
using Cartesian coordinates;
small amplitude and short wavelength perturbations
will also always be imposed.
The wavelength of the short-wavelength ``noise''
which grows depends on the resolution of the calculation.
To eliminate
the possibility of this effect in the current calculations, we also
performed the calculations with higher and lower resolution in the
$x$ direction; specifically, twice (384) and a half (96) the
number grid points than in our standard calculation.  Figure
\ref{resolution} shows the results of higher (right panels) and lower
(left panels) resolution calculations.  As a reference, the results
of the standard resolution calculation are also shown (center).
The results of the three different ways for imposing perturbations are
shown (top : $t=460$ kyr, 50\% peturbation in density;
middle : $t=440$ kyr, 30\% peturbation in photon number flux during
102-110 kyr; bottom : $t=460$ kyr, surface perturbation of
amplitude $9\times 10^{-3}$ pc).  The pillars appear and grow at
all resolutions, differing only in fine-scale structure at the
cloud surfaces.  The separation of the ionization front is also
observed at the early phase of the dynamics in each case.
Thus, we conclude that the pillar appearance is not a 
resolution-dependent effect.

\subsection{Discussion}
That we produce pillars
is purely the result of our periodic boundary conditions.
Unlike the pillars in our simulation,
the Eagle pillars are not co-planar
and do not share the same inclination angle (\citet{Pound98}).
They are effectively three isolated pillars, rather than a single
structure. Pillars in other HII regions are also typically isolated.
However, our main aim in this paper has been to describe the
instability of accelerating IF,
so we have assumed
initially a single wavelength and periodic boundary
conditions.
We have chosen only one set of free parameters---the
wavelength and number density, initial magnetic pressure of the
initial cloud, incident photon number flux at the boundary, and
thickness of the cloud---although these specific values match those
of Eagle Nebula reasonably well.

The main key to this instability is the mismatch of perturbation
modes between the IF and density in the HII region.  The separation
of the IF from the cloud is the trigger for this mismatch.
Thus, it should still be possible to create an isolated pillar in the
HII region under conditions where a single
wavelength grows and becomes a pillar.
Different wavelengths and multiwavelength
cases should be investigated, however,
since the phenomenon relies on nonlinear dynamics.

The effect of varying the magnetic field pressure should also be
investigated, although the magnetic pressure is introduced just to
prevent too high a number density from forming
and does not affect the physical
mechanism of the growth of the perturbation.  An MHD calculation
is one avenue for improvement over the current models.

In this study we have adopted the
on-the-spot approximation which ignores
the recombination of the ionized hydrogen to the ground state,
assuming that the photon emitted by this recombination is absorbed
locally because its energy is high enough to ionize another
hydrogen atom.  However, this approximation is not sufficient for
Cartesian calculations when the directions of IF propagation and
of the incident light rays become close to parallel.
In such cases, irradiation by diffuse photons is the main source of
ionization of the sides of the pillars.  We expect to include
this diffused photon transport in future calculations.

\section{CONCLUSION}
\label{conc}
We have investigated hydrodynamic instability of the accelerating
IF including detailed energy accounting.
Three different ways of imposing the initial perturbation
were adopted: (1) perturbation
in density: (2) perturbation in photon number flux: and (3) surface
perturbation.

We observe no large growth when the incident perturbation is small.
This stabilization is the same as that observed in Paper I,
and identical to the stabilization mechanism theoretically found by
\citet{Axford64} for the non-accelerating IF.
The strong absorption of incident photons
causes decreasing ablation pressure around the cavity,
which effectively smooths out small amplitude perturbations.

In contrast, large growth of the second harmonic of the imposed
perturbation is observed when the initial perturbation is large,
namely, the ratio of the initial amplitude to wavelength
is greater than 0.02.
A slight separation of the IF from the cloud surface triggers
the appearance of a warm HI region and the appearance of 
a second harmonic of the imposed perturbation.
The separation of the IF from the cloud
surface happens when all the incident photons traveling
toward this region are
absorbed by the recombined neutral hydrogen accumulated near the
bubble vertex.
The gas at this cloud surface at location begins to expand in the $y$
direction during the separation because the pressure in the warm
HI region is less than that in other regions.  Although the second
harmonic is imprinted on the cloud surface, the density perturbation
and incident
photon flux contours in the HII region are still single mode.
The mismatch between modes at the cloud surface
and in the density in the HII region
above the cloud surface prevents
stabilization and cause nonlinear dynamics of the
molecular cloud.  In the later phase of the dynamics, large
growth of the perturbation is observed.

This new result for the IF instability does not depend on the method
by which the initial perturbation is imposed.  We also show that it
is independent of the resolution of the calculation.  The velocity
gradient in the model pillar in the later phase of the dynamics is
in good agreement with that observed in the Eagle Nebula.

\acknowledgments

We would like to thank to Takashi Hosokawa, Mark Wolfire, Nobuhiko
Izumi, and Robin Williams for useful discussions and comments.
We acknowledge the anonymous referee for helpful comments
which improved this manuscript.
This work was performed under
the auspices of the U.S.  Department of Energy by the Lawrence
Livermore National Laboratory under Contract No. W-7405-ENG-48 and
with support from NASA Grant NRA 00-01-ATP-059 and from National
Science Foundation under Grant No. AST-0228974.

\clearpage

\end{document}